# A Polynomial-Time Deterministic Approach to the Traveling Salesperson Problem

Ali Jazayeri[1] and Hiroki Sayama[2]

[1]Department of Information Science, Drexel University, USA
[2]Center for Collective Dynamics of Complex Systems / Department of Systems Science and Industrial Engineering, Binghamton University, USA

**Abstract**

We propose a new polynomial-time deterministic algorithm that produces an approximated solution for the traveling salesperson problem. The proposed algorithm ranks cities based on their priorities calculated using a power function of means and standard deviations of their distances from other cities and then connects the cities to their neighbors in the order of their priorities. When connecting a city, a neighbor is selected based on their neighbors' priorities calculated as another power function that additionally includes their distance from the focal city to be connected. This repeats until all the cities are connected into a single loop. The time complexity of the proposed algorithm is $O(n^2)$, where $n$ is the number of cities. Numerical evaluation shows that, despite its simplicity, the proposed algorithm produces shorter tours with less time complexity than other conventional tour construction heuristics. The proposed algorithm can be used by itself or as an initial tour generator for other more complex heuristic optimization algorithms.

## Introduction

The traveling salesperson problem (TSP) has been broadly discussed in different contexts. The problem can be expressed as finding the shortest tour for a salesperson through $n$ cities by starting from his/her home city, visiting the other $n-1$ cities and returning back to the first city, visiting each city exactly once.

At least one optimum solution is always available, by counting all the possible permutations of $n$ cities. To find the optimum solution, the length of all possible $(n-1)!/2$ different routes should be computed. However, the number of routes in this case grows exponentially. Therefore, to avoid the considerable computational time, many algorithms have been proposed which, although they do not guarantee optimality, provide an approximate solution close to the optimal solution. Approaches used in these algorithms are diverse in nature, such as minimum spanning trees (Held and Karp 1970), genetic algorithms (Grefenstette, Gopal et al. 1985), evolutionary approaches (Fogel 1988), Markov chain Monte Carlo procedures (Martin, Otto et al. 1991), reinforcement learning (Dorigo and Gambardella 1995), and even unconventional computational methods (Jones and Adamatzky 2014).

Based on Papadimitriou (1977), the TSP can be classified as an NP-complete problem and therefore finding a polynomial-time algorithm that can return the optimal solution is very improbable. In this paper, a new polynomial-time deterministic approach is proposed, which can produce a solution with a small deviance from the true optimal solution with time complexity $O(n^2)$. In the next section, the proposed algorithm will be discussed. In the experimental findings section, the tours for 25 sets of cities from TSPLIB (with number of cities ranging from ~50 to ~1100) and 45 random Euclidean instances are used, and the time complexity and results of the proposed algorithm are compared with those obtained by other conventional tour construction heuristics. Finally, the paper concludes with some suggestions for further improvement.

**Proposed Algorithm**

The proposed algorithm consists of two main steps. It uses two different power functions in each main step to connect cities to their neighbors. The first power function is used to rank cities by considering the means and the standard deviations of their distances from all other cities and to connect those cities to their neighbors in the order of their priorities. When connecting a city, a neighbor is selected based on the neighbors' priorities calculated by the second power function that additionally includes their distance from the focal city to be connected. This repeats until all the cities have at least one connection in the first step, and exactly two connections in the second step. Through these two steps, the algorithm generates a single loop that connects all the cities. The whole algorithm is repeated for several different exponent values used in the power functions, and then the minimum total tour length is selected.

The ranking of cities using the first power function is based on the following intuitive fact: if distances between a specific city and all other cities are relatively equal, it does not matter much when that city is being connected, and therefore its priority over other cities should be low. Or, in a more formal language, if the standard deviation of distances of a city from all other cities is small, it has less priority to be connected to other cities, because its late connection would not have any significantly negative consequence on the length of the final tour. On the contrary, if a city has a fairly large amount of standard deviation of distances from other cities, then that city should have a high priority because if we keep the city unconnected while connecting its close neighbors to other cities, a huge negative consequence may arise.

In addition to the standard deviation of distances, the mean of distances should also be considered. This is because keeping a city that is far from other cities unconnected may eventually lead to a negative consequence at a later stage of route construction. Therefore, in this paper, we use the following power function to determine the priority of city $i$:

$$p_i = \mu_i^\alpha \sigma_i^\beta \qquad (1)$$

Here, $\mu_i$ and $\sigma_i$ are the mean and standard deviation of distances between city $i$ and the other cities. Due to the unknown exact influences of these two statistics, we include two exponent parameters, $\alpha$ and $\beta$, and consider several exponent values when calculating the priorities of cities.

Once the priorities of cities are calculated using Equation (1), the city with the highest priority whose number of neighbors is less than the corresponding step is chosen to select another city to connect. The neighbor city to be connected should be as close as possible to the focal city. Meanwhile, among potential neighbors, priority should be given to those cities whose late connection could cause greater negative consequences, based on the same argument discussed above. These two factors are combined in the following second power function:

$$c_j = \frac{\mu_j^\delta \sigma_j^\varepsilon}{d_j^{i\gamma}} \qquad (2)$$

Here, $c_j$ is the priority of neighbor city $j$ to be connected to city $i$, $d_j^i$ is the distance between city $i$ and city $j$, and $\mu_j$ and $\sigma_j$ are the mean and standard deviation of distances between city $j$ and all the other cities, respectively. Again, because of unknown exact influences of these three statistics, we introduce the three exponent parameters, $\gamma$, $\delta$ and $\varepsilon$, and their values are explored in the proposed algorithm.

The proposed algorithm proceeds as follows (also see Fig. 1):

- In the first main step, cities are processed in the order of the ranking calculated using Equation (1). Each city is connected to a neighbor city with the highest priority given by Equation (2), under the conditions that the degree of the focal city before connection should be zero and that the degree of the neighbor city should be less than two. This process is repeated for all cities. Once this first main step is complete, it is guaranteed that all the cities have at least one neighbor. Some may have two, as a result of being selected as the neighbor for other cities.
- In the second main step, Equation (1) is calculated again by using cities' mean and standard deviation of distances from all other cities. Then the cities that have just one neighbor are connected to another neighbor with the highest priority given by Equation (2), under the conditions that the connection of the neighbor city to the focal city will not create a cycle (unless this connection is the very last one to form a tour) and that the degree of the neighbor city is less than two before connection. This process is repeated for all cities. Once this second main step is complete, a single looped tour is generated.

Figure 1 shows the outline of the proposed algorithm.

➤ **Input:**

- o Set of cities and their distances from each other
- o Value sets for exponents of power functions used for ranking cities
- ➢ Shortest = infinity
- ➢ MeanList = list of means of distances from each city to all other cities
- ➢ STDList = list of standard deviations of distances from each city to all other cities
- ➢ The initial value of degree of each city (number of neighbors) is zero
- ➢ **For** different combinations of exponent values
  - o E = 0
  - o OtherEndList = [1 … n]
  - o **For** Step = 1, 2
    - ListOne = []
    - **For** City in 1 … n
      - **If** degree(City) < 2
        - Calculate Eq. (1) using MeanList[City], STDList[City] and the current exponent values
        - Append (City, Eq. (1) Result) to ListOne
    - **While** ListOne != []
      - City = the city with largest 2nd entry value in ListOne
      - Remove City from ListOne
      - **If** degree(City) < Step
        - ListTwo = []
        - **For** Neighbor in 1 … n except for City
          - **If** degree(Neighbor) < 2
            - **If** OtherEndList[City] != Neighbor or E == n − 1
              - o Calculate Eq. (2) using MeanList[Neighbor], STDList[Neighbor] and the current exponent values
              - o Append (Neighbor, Eq. (2) Result) to ListTwo
        - Neighbor = the city with the largest 2nd entry value in ListTwo
        - OtherEndList[OtherEndList[City]],OtherEndList[OtherEndList[Neighbor]] = OtherEndList[Neighbor], OtherEndList[City]
        - Connect City and Neighbor
        - E = E + 1
  - o **If** final tour length < Shortest
    - Shortest = final tour length
    - Remember used exponent values and connections among cities
- ➢ **Output:**
  - o Shortest
  - o Used exponent values
  - o Connections among cities

Figure 1: The pseudo-code for the proposed algorithm

Here we show that the computational complexity of the proposed algorithm is $O(n^2)$, where *n* is the number of cities. The algorithm includes two main steps. In each main step, $n - 1$ neighbor cities are evaluated for each city. Therefore, there are $n(n - 1)$ different processes for each step, which results in computational complexity $O(n^2)$. Finding the maximum values is performed in computational time $O(n)$, which does not change the total computational complexity. By creating and updating a list of other ends for cities, the time complexity of checking if there is a path between two cities can be reduced to $O(1)$ for each iteration. Repeating the whole algorithm for several combinations of exponent values multiplies computational complexity only by a constant factor, as long as the number of combinations is finite. Therefore, the computational complexity of the proposed algorithm is $O(n^2)$.

As an illustrative example, the proposed algorithm was applied to the 48 US capital cities (att48) and eil76 sets (from TSPLIB (Reinelt 1991)). The parameters for Equations (1) and (2) were selected from an exponent value set {0, 0.5, 1}. The best results were obtained with ($\alpha = 0.5$, $\beta = 1$, $\gamma = 1$, $\delta = 0.5$ and $\varepsilon = 0$) for att48 set and with ($\alpha = 0.5$, $\beta = 0$, $\gamma = 0.5$, $\delta = 0.5$ and $\varepsilon = 0.5$) for eil76 set. Figure 3 shows the results with these optimal exponents in more detail.

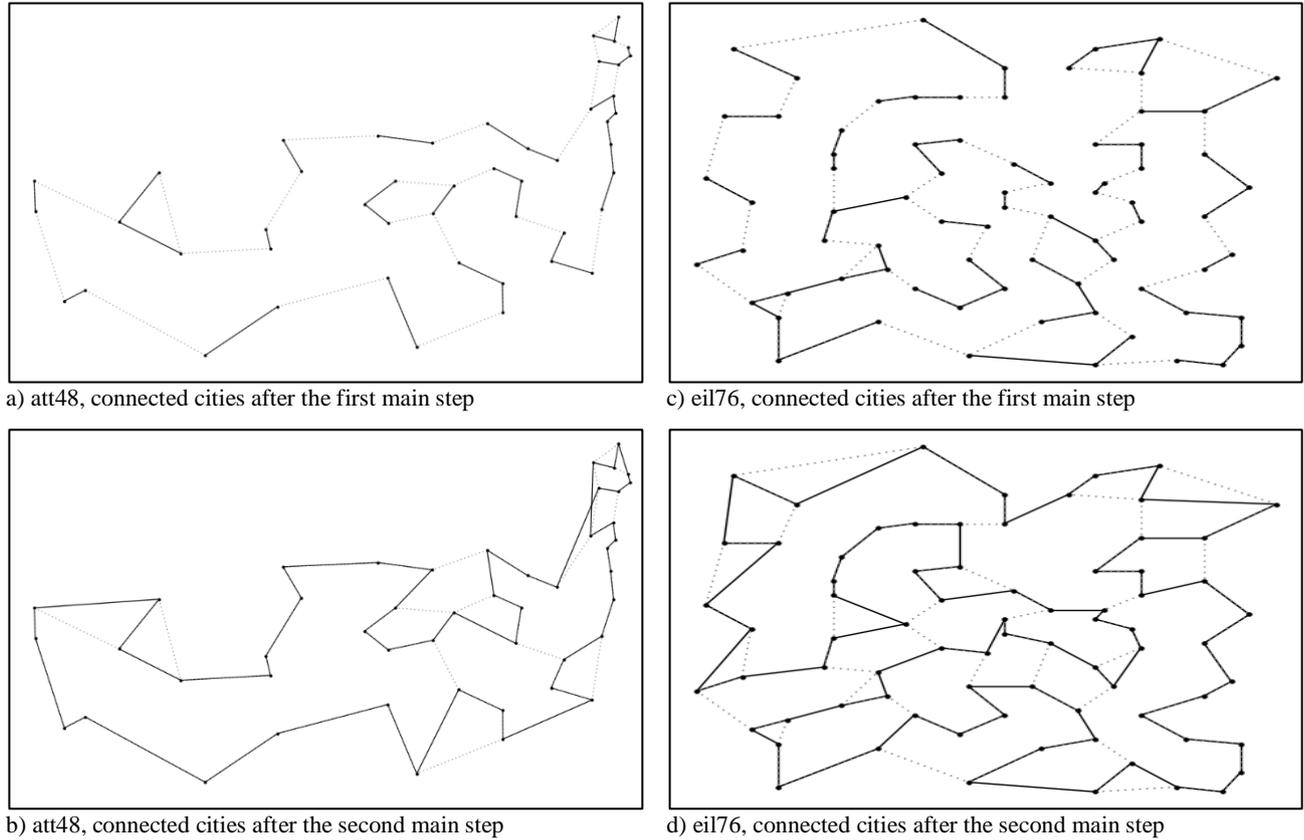

a) att48, connected cities after the first main step

c) eil76, connected cities after the first main step

b) att48, connected cities after the second main step

d) eil76, connected cities after the second main step

Figure 3: The proposed algorithm applied to the TSP problems with 48 US capital cities (att48 set) and eil76 set. **a & c**. Connected cities after the first main step. **b & d**. Connected cities after the second main step. In each figure, the connections obtained by the proposed algorithm are shown with solid lines (final tour length: 34839 and 565 for att48 and eil76 sets respectively), while the optimal tour is shown with dotted lines (true optimal tour length: 33,523 and 538 for att48 and eil76 sets respectively). The differences between the two routes lead to 3.93% and 5.02% deviations from the optimal solutions respectively for att48 and eil76.

## Experiments

The proposed algorithm was applied to 25 different sets of cities from TSPLIB and 45 random Euclidean instances. For comparing the performance of the proposed algorithm, four conventional tour construction algorithms, Nearest Neighbor, Greedy, Clarke-Wright, and Christofides were implemented (as they are discussed in the Johnson and McGeoch (1997) study) for both the TSPLIB and the 45 random Euclidean instances. For 45 random Euclidean instances, three numbers of cities were considered (100, 316 and 1000). For each number of cities, 15 random Euclidean instances were generated. The results of this algorithm for TSPLIB instances were compared to their corresponding best-known solutions in Table 1. As seen in the table, the best values for exponents of Equations (1) and (2) varied for different sets, indicating that optimal exponent values would depend on the topology of the network of cities and their spatial distances.

*Table 1: Performances of the proposed algorithm on 25 different sets of cities*

| Set Name | # of Cities | Best Route Length | Obtained Route Length | Error (%) | α | β | γ | δ | ε |
|---|---|---|---|---|---|---|---|---|---|
| att48 | 48 | 33523 | 34839 | 3.93% | 0.5 | 1 | 1 | 0.5 | 0 |
| eil51 | 51 | 426 | 453 | 6.44% | 0 | 0.5 | 1 | 0 | 0.5 |
| berlin52 | 52 | 7542 | 8023 | 6.38% | 0 | 0 | 0.5 | 1 | 0 |
| eil76 | 76 | 538 | 565 | 5.02% | 0.5 | 0 | 0.5 | 0.5 | 0.5 |
| kroA100 | 100 | 21282 | 22470 | 5.58% | 1 | 0.5 | 0.5 | 0.5 | 0 |
| kroB100 | 100 | 22141 | 24222 | 9.40% | 0.5 | 0 | 0.5 | 1 | 0 |
| kroC100 | 100 | 20749 | 22199 | 6.99% | 1 | 0.5 | 1 | 0.5 | 1 |
| kroD100 | 100 | 21294 | 22493 | 5.63% | 0.5 | 0 | 0.5 | 1 | 0 |
| kroE100 | 100 | 22068 | 23631 | 7.08% | 0.5 | 0 | 0.5 | 0 | 0 |
| lin105 | 105 | 14379 | 15132 | 5.23% | 0.5 | 0 | 0.5 | 1 | 0 |
| pr107 | 107 | 44303 | 50948 | 15.00% | 0.5 | 0 | 0.5 | 0 | 0 |
| bier127 | 127 | 118282 | 122461 | 3.53% | 0.5 | 1 | 1 | 0.5 | 0.5 |
| ch130 | 130 | 6110 | 6520 | 6.71% | 0 | 0.5 | 1 | 0.5 | 0 |
| ch150 | 150 | 6528 | 7015 | 7.47% | 0 | 0.5 | 0.5 | 0.5 | 0.5 |
| kroA150 | 150 | 26524 | 28572 | 7.72% | 0.5 | 0.5 | 0.5 | 0.5 | 0.5 |
| kroB150 | 150 | 26130 | 28184 | 7.86% | 0.5 | 1 | 0.5 | 0 | 1 |
| d198 | 198 | 15780 | 17081 | 8.24% | 0.5 | 1 | 0.5 | 0 | 0.5 |
| kroA200 | 200 | 29368 | 31473 | 7.17% | 1 | 0.5 | 0.5 | 0.5 | 0 |
| gil262 | 262 | 2378 | 2572 | 8.16% | 0.5 | 0 | 1 | 1 | 0.5 |
| lin318 | 318 | 42029 | 45850 | 9.09% | 1 | 0.5 | 0.5 | 0.5 | 0 |
| d493 | 493 | 35002 | 37653 | 7.57% | 1 | 0.5 | 1 | 1 | 0.5 |
| dsj1000 | 1000 | 18660188 | 20815781 | 11.55% | 0.5 | 0 | 1 | 0.5 | 0 |
| pr1002 | 1002 | 259045 | 284505 | 9.83% | 0.5 | 0 | 0.5 | 0.5 | 1 |
| u1060 | 1060 | 224094 | 248005 | 10.67% | 0.5 | 0 | 0.5 | 1 | 0 |
| vm1084 | 1084 | 239297 | 265738 | 11.05% | 0.5 | 0 | 0.5 | 0 | 0 |

The average error for these 25 sets of cities is 7.73%. In comparison with other more conventional tour construction techniques summarized by Johnson and McGeoch (1997), the proposed algorithm exhibits a smaller average error with equal or less computational complexity, as the performance results for TSPLIB and random Euclidean instances in comparison with the performance of the four conventional tour construction are shown in Tables 2 and 3 respectively. For random Euclidean instances, the Held-Karp bound were computed based on the proposed formula by Johnson and McGeoch (1997).

*Table 2: performances of four tour construction heuristics in comparison with the proposed algorithm for TSPLIB instances, %error is calculated based on the best-known route length of Table 1*

| Set Name | # of Cities | %Error | | | | |
|---|---|---|---|---|---|---|
| | | Nearest Neighbor | Greedy | Clarke-Wright | Christofides | Proposed algorithm |
| att48 | 48 | 20.89% | 19.80% | 4.45% | 20.29% | 3.93% |
| eil51 | 51 | 20.57% | 13.03% | 2.62% | 17.33% | 6.44% |
| berlin52 | 52 | 19.08% | 31.98% | 9.93% | 12.16% | 6.38% |
| eil76 | 76 | 32.34% | 14.71% | 6.73% | 17.20% | 5.02% |
| kroA100 | 100 | 26.19% | 13.70% | 8.30% | 18.53% | 5.58% |
| kroB100 | 100 | 31.68% | 16.59% | 10.17% | 9.19% | 9.40% |
| kroC100 | 100 | 26.88% | 12.94% | 8.52% | 13.68% | 6.99% |
| kroD100 | 100 | 26.56% | 14.82% | 7.55% | 13.47% | 5.63% |
| kroE100 | 100 | 25.01% | 12.59% | 8.86% | 9.31% | 7.08% |
| lin105 | 105 | 41.61% | 16.64% | 7.55% | 26.26% | 5.23% |
| pr107 | 107 | 5.36% | 6.60% | 10.96% | 9.70% | 15.00% |
| bier127 | 127 | 14.77% | 19.31% | 4.74% | 11.16% | 3.53% |
| ch130 | 130 | 23.98% | 28.40% | 11.58% | 13.15% | 6.71% |
| ch150 | 150 | 25.53% | 18.61% | 6.81% | 9.11% | 7.47% |

| | | | | | |
|---|---|---|---|---|---|
| kroA150 | 150 | 26.71% | 20.24% | 7.43% | 16.13% | 7.72% |
| kroB150 | 150 | 25.62% | 20.25% | 6.52% | 22.54% | 7.86% |
| d198 | 198 | 18.00% | 22.20% | 11.71% | 14.57% | 8.24% |
| kroA200 | 200 | 21.90% | 17.83% | 9.49% | 13.43% | 7.17% |
| gil262 | 262 | 36.31% | 13.21% | 11.20% | 13.19% | 8.16% |
| lin318 | 318 | 28.56% | 18.75% | 8.22% | 17.40% | 9.09% |
| d493 | 493 | 24.70% | 16.12% | 10.85% | 15.87% | 7.57% |
| dsj1000 | 1000 | 32.00% | 16.32% | 9.46% | 15.39% | 11.55% |
| pr1002 | 1002 | 21.85% | 21.15% | 10.39% | 12.52% | 9.83% |
| u1060 | 1060 | 25.68% | 21.07% | 10.56% | 13.14% | 10.67% |
| vm1084 | 1084 | 25.98% | 19.60% | 8.97% | 12.33% | 11.05% |
| **Mean (SD):** | | **25.11% (6.95%)** | **17.86% (5.06%)** | **8.54% (2.29%)** | **14.68% (4.11%)** | **7.73% (2.51%)** |

*Table 3: Computational complexities of four conventional tour construction heuristics and their performances in comparison with the proposed algorithm for random Euclidean instances, %error is calculated based on the Held-Karp bound*

| Method | Computational Complexity | % Average Error | | | Mean (SD) |
|---|---|---|---|---|---|
| | | $n = 100$ | $n = 316$ | $n = 1000$ | |
| **Nearest Neighbor** | $O(n^2)$ | 23.17% | 26.52% | 24.38% | **24.69% (4.73%)** |
| **Greedy** | $O(n^2 \log n)$ | 17.00% | 17.89% | 17.21% | **17.36% (4.04%)** |
| **Clarke-Wright** | $O(n^2 \log n)$ | 9.75% | 10.13% | 10.88% | **10.25% (2.65%)** |
| **Christofides** | $O(n^3)$ | 14.30% | 15.89% | 14.67% | **14.95% (2.81%)** |
| **Proposed algorithm** | $O(n^2)$ | 6.34% | 7.67% | 8.81% | **7.61% (1.95%)** |

Reinelt (1994) compared different tour construction algorithms and discussed that, considering the absolute and relative quality of tour lengths and also taking into account the required running time, the Clarke-Wright algorithm (called Savings heuristics by Reinelt) provides the best solution. Table 2 shows that our proposed algorithm performed better than those algorithms, including Clarke-Wright

We should note that those other algorithms listed in Table 2 are known to produce only moderate quality solutions by themselves. To reduce the length of solutions, a number of improvement heuristics have been applied to them, such as node/edge insertion, crossing elimination, Lin-Kernighan type heuristics, and other heuristic approaches to prevent the stochastic search from getting stuck in local minima. However, those improvements would require significantly greater computational resource (Reinelt 1994). In contrast, the proposed algorithm can provide better solutions deterministically and its computational complexity is low.

## Conclusion

In this paper, we have proposed a new algorithm for finding solutions for the traveling salesperson problem. The algorithm is composed of two main steps. The main contribution of the proposed algorithm is the introduction of this intuitive assumption that we can rank cities according to their distances to other cities and their standard deviations , both to select their neighbors and for being selected by their neighbors. The algorithm starts with connecting cities with higher priorities to other cities with higher priorities as neighbors. Furthermore, computational complexity, deterministic nature, and simplicity of execution can be considered as the main strengths of the proposed algorithm.

Here we assumed only three values, $\{0, 0.5, 1\}$, for exponents in the power functions used for ranking cities. Other values, even negative ones, may be considered to improve the performance of the algorithm. Another way to improve the final results is to consider only closer cities for each city; excluding farthest cities might result in improvement of priority lists. In addition, the output of this algorithm can be considered as an input to other heuristic algorithms. One important further avenue of research is to explore different forms of functions to rank cities by using other kinds of statistics.